\documentclass[pra,aps,showpacs,twocolumn,floatfix]{revtex4}
\usepackage{graphicx}
\usepackage{array}
\usepackage{color}
\usepackage{amsmath}
\usepackage{amsxtra}
\usepackage{amstext}
\usepackage{amssymb}
\usepackage{latexsym}
\usepackage{dsfont}

\begin{document}

\title{Squeezed cooling of mechanical motion beyond the resolved-sideband limit}
\author{Lin Zhang}
\author{Cheng Yang}
\affiliation{School of Physics and Information Technology, Shaanxi Normal University,
Xi'an 710061, P. R. China}
\author{Weiping Zhang}
\affiliation{Department of Physics and Astronomy, Shanghai Jiao Tong University,
Shanghai 200240, P. R. China}

\begin{abstract}
Cavity optomechanics provides a unique platform for controlling micromechanical
systems by means of optical fields that crosses the classical-quantum boundary to achieve
solid foundations for quantum technologies. Currently, optomechanical resonators have become
promising candidates for the development of precisely controlled nano-motors, ultrasensitive
sensors and robust quantum information processors. For all these applications, a crucial
requirement is to cool the mechanical resonators down to their quantum ground states.
In this paper, we present a novel cooling  scheme to further cool a micromechanical
resonator via the noise squeezing effect.  One quadrature in such a resonator can be
squeezed to induce enhanced fluctuation in the other, ``heated'' quadrature, which can
then be used to cool the mechanical motion via conventional optomechanical coupling. Our
theoretical analysis and numerical calculations demonstrate that this squeeze-and-cool
mechanism offers a quick technique for deeply cooling a macroscopic mechanical resonator
to an unprecedented temperature region below the zero-point fluctuations.
\end{abstract}

\maketitle

\begin{figure}[htp]
\includegraphics[width=0.48 \textwidth]{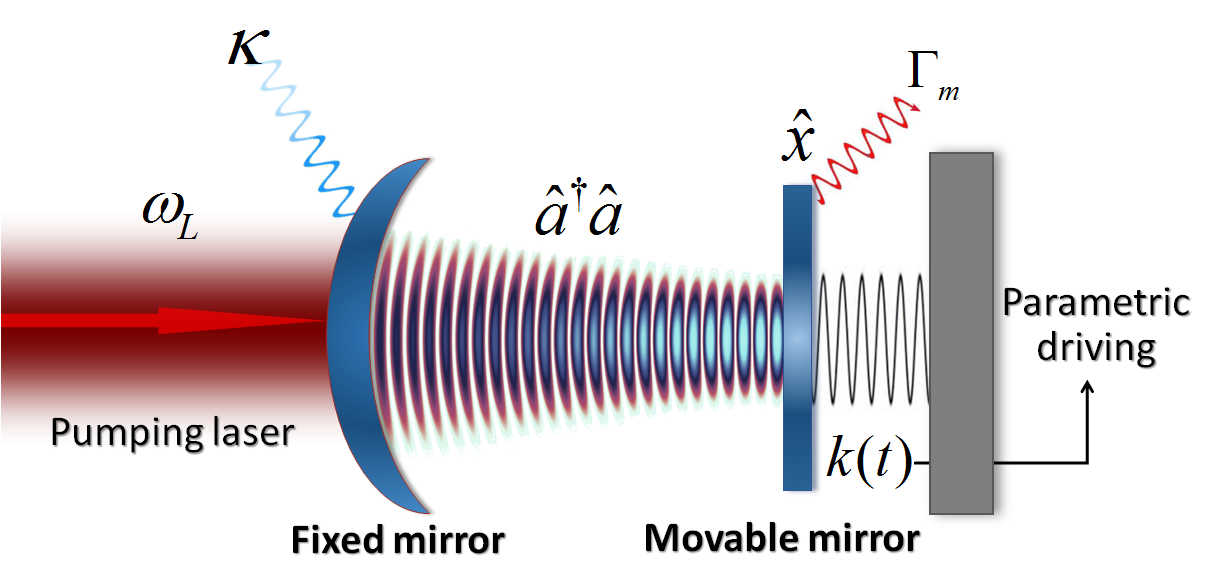}%
\caption{Schematic diagram of a hybrid optomechanical system with a parametric mirror
independently driven by the modulation of its spring constant $k(t)$ and coupled to a
cooling cavity mode.}
\label{figure1}
\end{figure}

Cavity optomechanics \cite{Aspelmeyer} concerns the strong interactions between
optical fields and mechanical oscillators that are derived from the mechanical
effects of photons. Radiation pressure is a major light-induced mechanical force
arising from the fact that a massless photon carries momentum. Linear or angular
momentum can be transferred from light fields to mechanical objects when photons
are absorbed or emitted. However, the mechanical effect of radiation pressure
on a macroscopic object is extremely weak \cite{Nichols}. Fortunately,
high-$Q$ optical cavities can resonantly enhance this optical force by trapping
high-intensity light within a very small volume, such as a typical linear
Fabry-P\'{e}rot cavity consisting of a heavy fixed mirror and a light movable mirror
attached to an elastic boundary (see Fig.\ref{figure1}) \cite{Aspelmeyer}. Since a
photon is reflected multiple times (e.g., $10^{6}$ \cite{Kleckner}) between the
two cavity mirrors before it decays, an intense cavity field builds up, resulting
in a large optical force on the movable mirror, which causes the mirror to vibrate
at a frequency ranging from kHz to GHz \cite{Vahala}. In recent decades, the
field of cavity optomechanics has witnessed rapid growth, and related research has
become increasingly important for both fundamental physics and applied technology \cite{MichaelMetcalfe}.

Efficient cooling of a massive mechanical resonator to its quantum ground state is
a prominent achievement of cavity optomechanics \cite{A.Schliesser,Gigan,Marquardt},
and a mechanical resonator with an average phonon occupation of $0.20\pm 0.02$ has recently
been achieved \cite{Peterson}. Cooling an oscillation mode to the ground state with
a noise below the standard quantum limit is a fundamental requirement for various optomechanical
applications, such as reliable nano-motors, high-precision sensors and robust quantum processors.
The basic physics of conventional optomechanical cooling is that the coupled cavity field introduces
additional friction to adaptively reduce the momentum of the resonator via an optical-spring effect
\cite{Braginsky,Arcizet}.
However, the quantum uncertainty principle prevents the complete halting of the resonator's motion
to access a temperature beyond the quantum fluctuations. In the attempt to achieve deeper
cooling towards quantum limit \cite{AAClerk},
the typical mechanism based on dynamical backaction loses efficacy due to the quantum backaction
limit, and various alternative cooling schemes within the resolved-sideband limit have
been proposed to achieve lower temperatures \cite{Teufel,Schliesser,Park,J.Chan}. Although a
recent cooling experiment using squeezed light have overcome the quantum backaction limit \cite{Clark},
all sideband cooling techniques are eventually limited by the single-phonon scattering balance \cite{Peterson},
which makes ground-state cooling below the one-phonon level more difficult due to a low final efficiency.

In this paper, we present an alternative technique for deep cooling via squeezing. The basic idea of
this technique is very similar to that of magnetic refrigeration. The confinement of magnetic
dipoles in one direction by an external magnetic field will drive fluctuations in the spatial
degrees of freedom into the momentum channel,
thereby effectively improving the temperature of the refrigerant. This thermal squeezing effect
can be essentially understood in terms of an oscillator entering a tighter potential
with a more confined position, causing its momentum to increase. Thus, the problem becomes one
of cooling a hotter refrigerant, which is easier than cooling a cold one. When the confinement
is finally removed, the motion fluctuations will return back to the spatial degrees of freedom,
and the overall temperature will decrease. Similarly, in the quantum domain, when one quadrature
of a mechanical resonator is squeezed, the other conjugate quadrature will be ``heated" by increased
quantum fluctuations due to the quantum uncertainty principle \cite{temperature}. For a given
optomechanical oscillator, the quantum fluctuations in the momentum quadrature $p$ can be increased by
quantum squeezing on the position quadrature $x$. Then, the ``hotter" quadrature, with its amplified
fluctuations, can be directly coupled to a blue-detuned cavity mode for conventional optomechanical cooling.

The quantum noise squeezing effect and its applications have been understood for approximately $50$ years
\cite{Caves,Yamamoto}, and noise squeezing on optomechanical resonators has also
been achieved using various schemes \cite{D.Rugar,Castellanos,Woolley,Wollman,Pirkkalainen}.
Because parametric squeezing is a robust technique applied in many quantum
and classical systems \cite{A.Szorkovszky,J-QLia,Szorkovszky,G.S.Agarwal}, we choose
parametric driving as our means of creating mechanical squeezing and theoretically study the cooling
scheme for an optomechanical oscillator that is squeezed by parametric amplification \cite{Huang}.
Our theoretical analysis shows that this squeezed resonator can be effectively cooled down to a low
temperature that is limited only by the degree of squeezing. With repeated cycles of squeezing and
cooling, the resonator can be deeply cooled to a temperature region that is inaccessible for
conventional cooling methods without noise squeezing.

The main schematic of our system, which consists of a high-$Q$ cavity and a movable mirror
(a mechanical oscillator), is shown in Fig.\ref{figure1}. The mechanical oscillator, with
a frequency of $\omega _{m}$ and an effective mass of $m_{eff}$, is separately controlled
by parametric driving on the spring constant $k(t)$ at a double frequency of $\omega _{m}$
and with a driving shift of $\delta$. Thus, the Hamiltonian of the driven parametric oscillator
(DPO) is \cite{A.Szorkovszky}
\begin{equation}
\hat{H}_{o}=\frac{\hat{p}^{2}}{2m_{eff}}+\frac{1}{2}\left[ k_{0}-k_{r}\sin
\left( \omega _{d}t+2\theta \right) \right] \hat{x}^{2},  \label{Ho}
\end{equation}%
where the free spring constant is $k_{0}=m_{eff}\omega _{m}^{2}$, the driving
frequency is $\omega _{d}=2(\omega _{m}-\delta )$, the driving amplitude is $k_{r}$,
and $\theta $ is the driving-induced phase shift. For a classical DPO, strong squeezing
of thermomechanical noise has been experimentally demonstrated \cite{D.Rugar}. To
highlight the quantum squeezing effect, the original Hamiltonian $\hat{H}_{o}$ in a
frame rotating at a frequency of $\omega_{m}-\delta$ can be written as \cite{A.Szorkovszky}
\begin{equation}
\hat{H}_{or}=\hbar \delta \hat{b}^{\dag }\hat{b}+i\frac{\hbar }{2}\left( \xi
^{\ast }\hat{b}^{2}-\xi \hat{b}^{\dag 2}\right) ,  \label{Hr}
\end{equation}%
where $\hat{b}$ ($\hat{b}^{\dag }$) is the phonon annihilation (creation) operator,
with an effective mechanical frequency of $\delta$. The second term of $\hat{H}_{or}$
takes the exact form of a squeezing operator, which can generate quantum noise squeezing
on the DPO. The squeeze parameter is $\xi =re^{-2i\theta }$, where the squeeze factor is $%
r=\omega _{m}k_{r}/4k_{0}$ and the phase shift $\theta $ adjusts the squeezing
directions of the coupled quadratures \cite{D.F.Walls}. Therefore, the hybrid system
depicted in Fig.\ref{figure1} can be described by the following Hamiltonian:
\begin{equation}
\label{Ht}
\hat{H}_{hyb}=\hat{H}_{or}+\hbar \delta _{c}\hat{a}^{\dag }\hat{a}-\hbar g%
\hat{a}^{\dag }\hat{a}(\hat{b}^{\dag }+\hat{b})+i\hbar (\eta \hat{a}^{\dag
}-\eta ^{\ast }\hat{a}).
\end{equation}%
The second term of $\hat{H}_{hyb}$ represents a cavity mode $\hat{a}$ with a detuning
of $\delta _{c}=\omega_{cav}-\omega _{L}$, where $\omega _{cav}$ and $\omega _{L}$ are
the frequencies of the cavity mode and the pumping laser, respectively. The third
term describes the optomechanical coupling between the cavity mode $\hat{a}$ and
the mechanical mode $\hat{b}$, which has a coupling strength of $g$, and the last
term represents the laser pumping with a strength of $\eta$ \cite{SM}.

Because the entire system is subjected to fluctuations originating from both the
external reservoirs and the internal quantum dynamics, the full motion of the system
can be described by the quantum Langevin equations of $\hat{H}_{hyb}$, which contain
noise terms for both the optical mode ($\hat{a}_{in}$) and the mechanical mode ($\hat{b}_{in}$)
as follows:
\begin{eqnarray}
\frac{d\hat{a}}{dt} &=&-\left( i\delta _{c}+\frac{\kappa }{2}\right) \hat{a}%
+ig\hat{a}(\hat{b}^{\dag }+\hat{b})+\eta +\sqrt{\kappa }\hat{a}_{in},
\label{at} \\
\frac{d\hat{b}}{dt} &=&-\left(i\delta+ \frac{\Gamma _{m}}{2}\right) \hat{b}%
+ig\hat{a}^{\dag }\hat{a}-\xi \hat{b}^{\dag }+\sqrt{\Gamma _{m}}\hat{b}_{in},
\label{bt}
\end{eqnarray}%
where $\kappa $ is the total decay rate of the cavity mode and $\Gamma _{m}$ is
the damping rate of the mechanical mode \cite{SM}. The noise operators
$\hat{a}_{in}$ and $\hat{b}_{in}$ represent the corresponding vacuum fluctuations,
which have the following statistical reservoir properties: $\langle \hat{a}_{in}(t)
\hat{a}_{in}^{\dag}(t^{\prime })\rangle =(n_{cav}^{th}+1)\delta (t-t^{\prime })$,
$\langle\hat{a}_{in}^{\dag }(t)\hat{a}_{in}(t^{\prime })\rangle =n_{cav}^{th}\delta
(t-t^{\prime })$, $\langle \hat{b}_{in}(t)\hat{b}_{in}^{\dag }(t^{\prime
})\rangle =(n_{m}^{th}+1)\delta (t-t^{\prime })$, and $\langle \hat{b}_{in}^{\dag
}(t)\hat{b}_{in}(t^{\prime })\rangle =n_{m}^{th}\delta (t-t^{\prime })$. Here, we
assume that the cavity field contains a number of thermal photons equal to $%
n_{cav}^{th}=[\exp (\hbar \omega _{cav}/k_{B}T)-1]^{-1}$ and
that the number of thermal phonons is given by $n_{m}^{th}=[\exp (\hbar \omega
_{m}/k_{B}T)-1]^{-1}$ \cite{Bowen}. 
\begin{figure}[htp]
\includegraphics[width=0.48\textwidth]{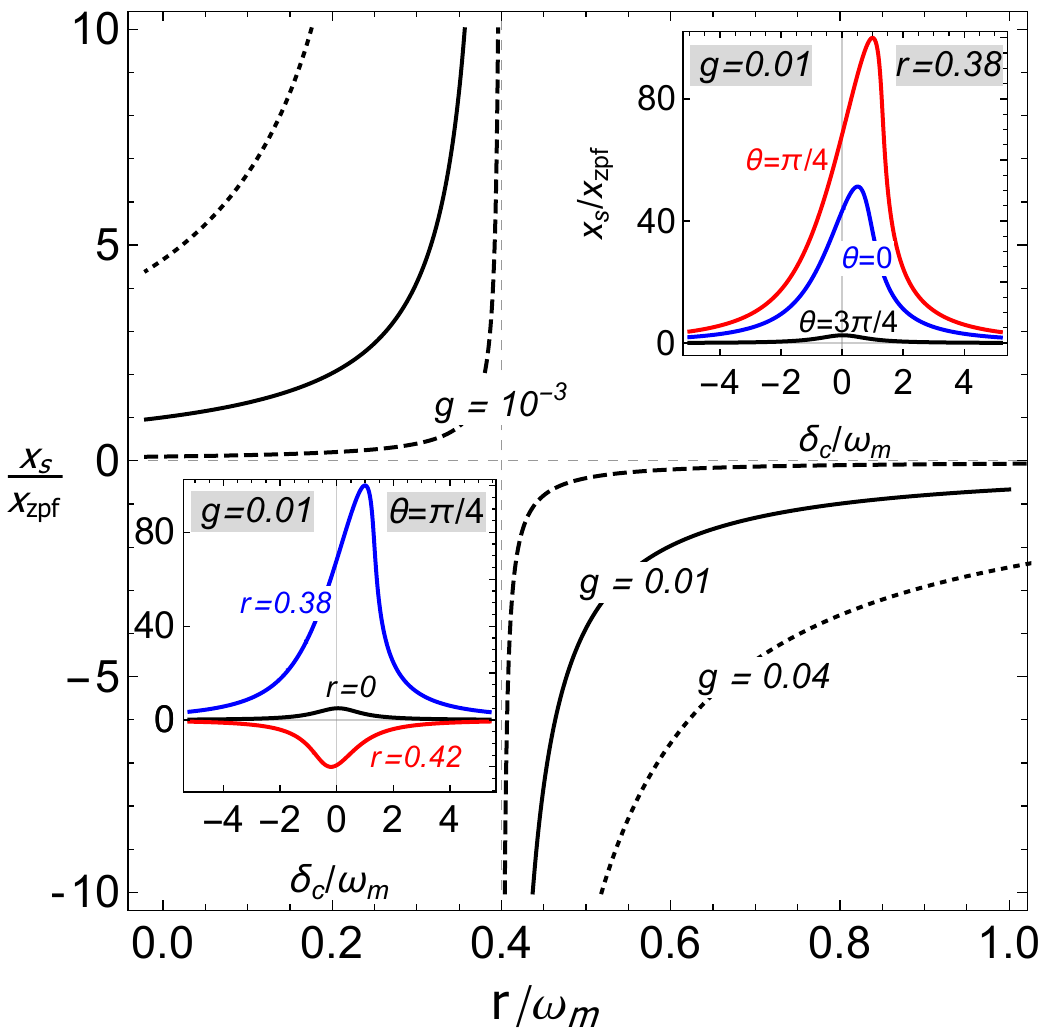}
\caption{The static displacements of the DPO modified by the
squeeze factor $r$ ($\protect\theta=\pi/4$) with different optomechanical coupling rates $g$.
Insets: The displacement responses of $x_s$ with respect to the pumping detuning $\protect\delta_c$
under different squeezing parameters. The other parameters are $\delta_c=2$, $\kappa =2$,
$\delta=0.4$, $\Gamma_m=10^{-3}$, and $\eta=10$, all scaled by $\omega_m$.}
\label{figure2}
\end{figure}

In the weak-pumping regime after cryogenic precooling, all dynamical quantities modulate
around their classical equilibrium states as follows: $\hat{a}=a_{s}+\delta \hat{a}$, $\hat{a}^{\dag
}=a_{s}^{\ast }+\delta \hat{a}^{\dag }$, $\hat{b}=b_{s}+\delta \hat{b}$, and $\hat{b}%
^{\dag }=b_{s}^{\ast }+\delta \hat{b}^{\dag }$. Thus, the equilibrium position of
the resonator can be determined using the following implicit equation:
\begin{equation}
x_{s}=\frac{2g\eta ^{2}\left( \delta +r\sin 2\theta \right) }{\left[ \left(
\kappa /2\right) ^{2}+(\delta_{c}-g x_{s})^{2}\right] \left[
\left( \Gamma _{m}/2\right) ^{2}+\delta ^{2}-r^{2}\right] },  \label{xs}
\end{equation}%
where $x_s=b_{s}+b_{s}^{\ast}$.
In Fig.\ref{figure2}, a squeezing-enhanced sensitive displacement of $x_s$ near
the critical squeezing point of $r \sim r_c=\sqrt{\delta^{2}+\Gamma_{m}^{2}/4}$
is identified, and it implies an efficient method of phonon cooling when $r>r_c$ \cite{SM}.
The nonlinear static responses of $x_s$ with respect to the pumping detuning $\delta_c$ also
sensitively depend on $r$ (lower inset) and $\theta$ (upper inset). To ensure reliable
performance, the stability requirement of the DPO imposes an upper bound on the squeezing
degree of $r$ such that $r<r_c$ for steady-state parametric squeezing \cite{Szorkovszky,SM}.
However, the instability beyond the squeezing bound can be overcome in two ways. One is to use a
red-detuned cooling laser to induce a positive damping rate $\Gamma_{opt}$ on the DPO (through
optical spring and damping effects), which can stabilize the steady-state squeezing over a large
squeezing region \cite{Schediwy}. The other reliable technique is to introduce quantum feedback
control \cite{Wilson,Vitali} over the dynamics of the DPO that extends beyond its steady-state
performance (e.g., locking onto a controlled self-sustained oscillation) and can work in the
deep squeezing regime \cite{A.Szorkovszky}.

For simplicity, we introduce the replacements $\delta \hat{a}\rightarrow \hat{a}$ and $\delta%
\hat{b}\rightarrow \hat{b}$ to obtain an effective linearized Hamiltonian of the system:
\begin{eqnarray}
\hat{H}_{eff} &=&\hbar \Delta_c \hat{a}^{\dag }\hat{a}+\hbar \delta \hat{b}%
^{\dag }\hat{b}-\hbar g(a_{s}^{\ast }\hat{a}+a_{s}\hat{a}^{\dag })(\hat{b}%
^{\dag }+\hat{b})  \notag \\
&&+i\frac{\hbar }{2}(\xi ^{\ast }\hat{b}^{2}-\xi \hat{b}^{\dag 2}),
\label{Heff}
\end{eqnarray}
where $\Delta_c$ is the position-shifted cavity detuning and is defined as $\Delta_c \equiv
\delta_{c}-g x_{s}$. Clearly, the last term of Eq.(\ref{Heff}) generates noise squeezing on
the mechanical mode $\hat{b}$ by means of the squeeze operator $\hat{S}(\xi)=\exp(%
\frac{1}{2}\xi^{\ast}\hat{b}^{2}-\frac{1}{2}\xi\hat{b}^{\dag2})$, where $\xi =re^{-2i\theta }$
controls the degree $r$ and the angle $\theta$ of the squeezing
on the coupled quadratures \cite{D.F.Walls}.

With the physical setup described above, our system can simultaneously apply mechanical squeezing
to one quadrature and perform cooling on the other. The squeezing-and-cooling effect can be
analyzed based on the phonon noise spectrum of the DPO.
The final number of phonons in the mechanical resonator is determined by
\begin{equation}
\bar{n}=\langle \hat{b}^{\dag }\left( t\right) \hat{b}\left( t\right)
\rangle =\frac{1}{2\pi}\int_{-\infty }^{\infty
}S_{n}\left( \omega \right) d\omega ,
\end{equation}%
where $S_{n}\left( \omega \right)$
is the phonon-number spectrum \cite{SM}, and this implies
an effective cooling temperature $T_{eff}$ of
\begin{equation}
T_{eff}(\bar{n})=\frac{\hbar \omega _{m}}{\ln \left( \frac{1}{\bar{n}}+1\right)}.
\end{equation}%
The above relation is derived from the detailed balance expression \cite{Marquardt,AAClerk,Bowen}, and
$\bar{n}$ is equal to the area underneath the spectral curve $S_n(\omega)$, implying that
$\lim_{\bar{n}\rightarrow 0}T_{eff}\rightarrow 0$.%
\begin{figure}[htp]
\includegraphics[width=0.48 \textwidth]{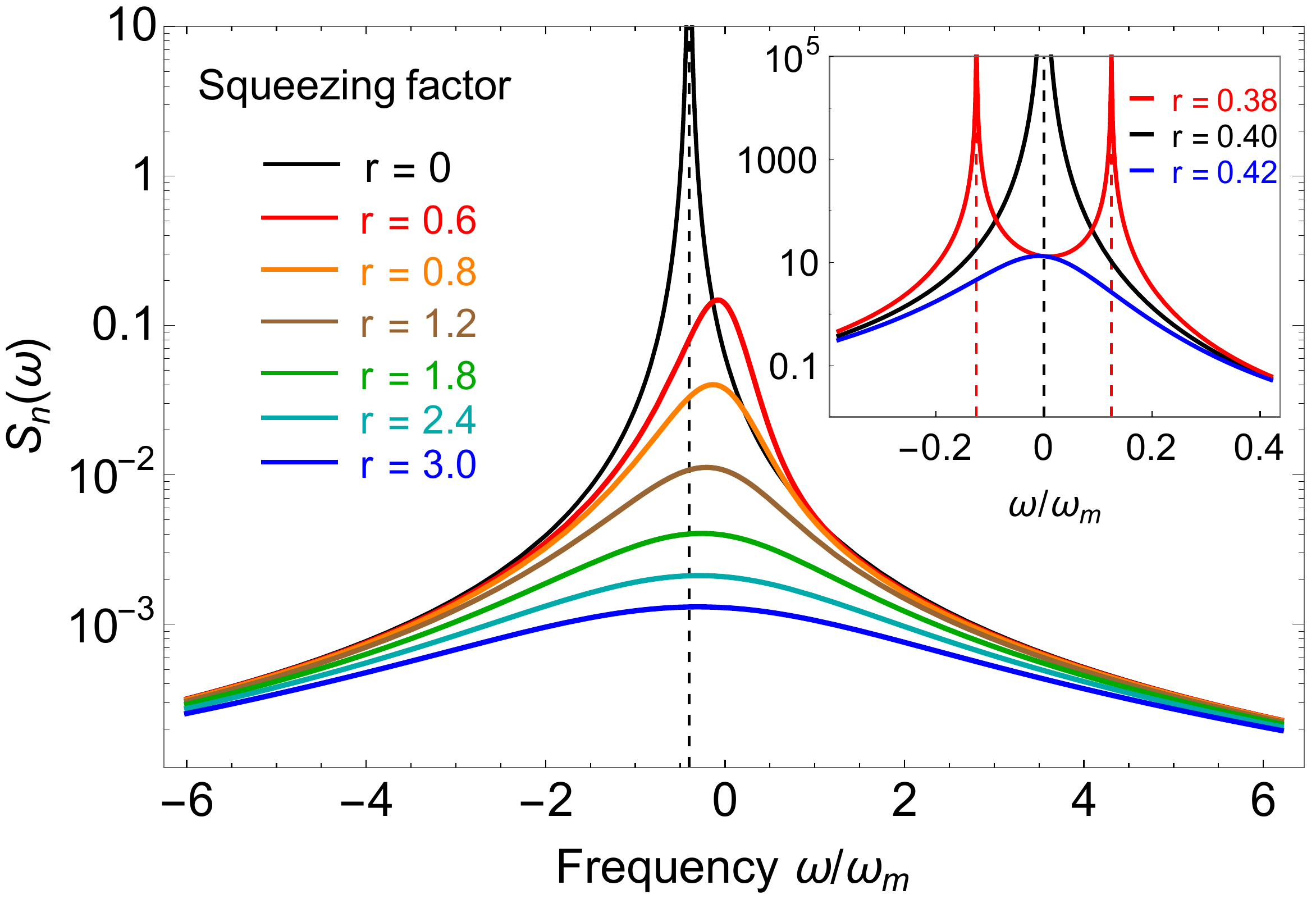}\newline
\caption{The phonon-number spectra of the mechanical
resonator for different squeeze factors $r>r_c$ (scaled by $\omega_m$) with a fixed
squeezing angle of $\theta=\pi/4$. Inset: phonon spectra for squeeze factors
$r$ near the critical squeezing factor $r_c\approx 0.4$. The effective mechanical
frequency is $\delta =0.4$ (indicated by the dashed vertical line), the optomechanical coupling
rate is $g=0.001$, the average thermal occupation is $n_{m}^{th}=10$, and the optical-mode
occupation is $n_{cav}^{th}=1$. The other parameters are $\delta _{c}=2$,
$\kappa=2$, $\Gamma _{m}=10^{-3}$, and $\eta=10$, all scaled by $\omega_m$.}
\label{figure3}
\end{figure}

In Fig.\ref{figure3}, we present the calculated phonon-number spectra $S_n(\omega)$ for different
squeeze factors $r$. The spectra exhibit a significant decrease in $T_{eff}$ with an increasing
squeeze factor $r$.
The inset figure reveals the critical squeeze factor $r_c$ and the marked squeezing-based cooling
that occurs when $r>r_c$ \cite{SM}. Here, we have chosen moderate parameters that
can be easily realized in optomechanical systems \cite{Clark}. A high-$Q$ resonator with
$\Gamma_{m}/\omega_m=0.001$ is used to support the squeezing performance, and a precooling process is applied that
achieves an average thermal occupation of $n^{th}_{m}=10$ to enhance the noise squeezing effect.
The cavity mode, with an occupation of $n^{th}_{cav}=1$, should be pumped by a far-red-detuned
low-power laser source. These undemanding conditions for squeezing-based cooling can be easily
fulfilled in a bad cavity ($\kappa/\omega_m=2$) beyond the resolved-sideband cooling limit and
in the weak-coupling regime ($g<\kappa$). The cooling mechanism takes effect in this regime because
the squeezed ``hotter" phonons with enhanced fluctuations are squeezed out and quickly taken
away by the coupled photons leaking from the bad cavity. Because a larger squeeze factor $r$
(the lower spectra shown in Fig.\ref{figure3}) will drive the system into an unstable state,
this technique is limited only by the squeeze factor $r/\omega _{m}=k_{r}/4k_{0}$ that is determined
by the relative modulation amplitude of the parametric resonator. Regardless, a small squeeze
factor of $r_c<r<2$ is still effective because we can use successive squeezing-and-cooling
cycles to achieve a lower temperature \cite{Vitali}.

\begin{figure}[htp]
\includegraphics[width=0.48 \textwidth]{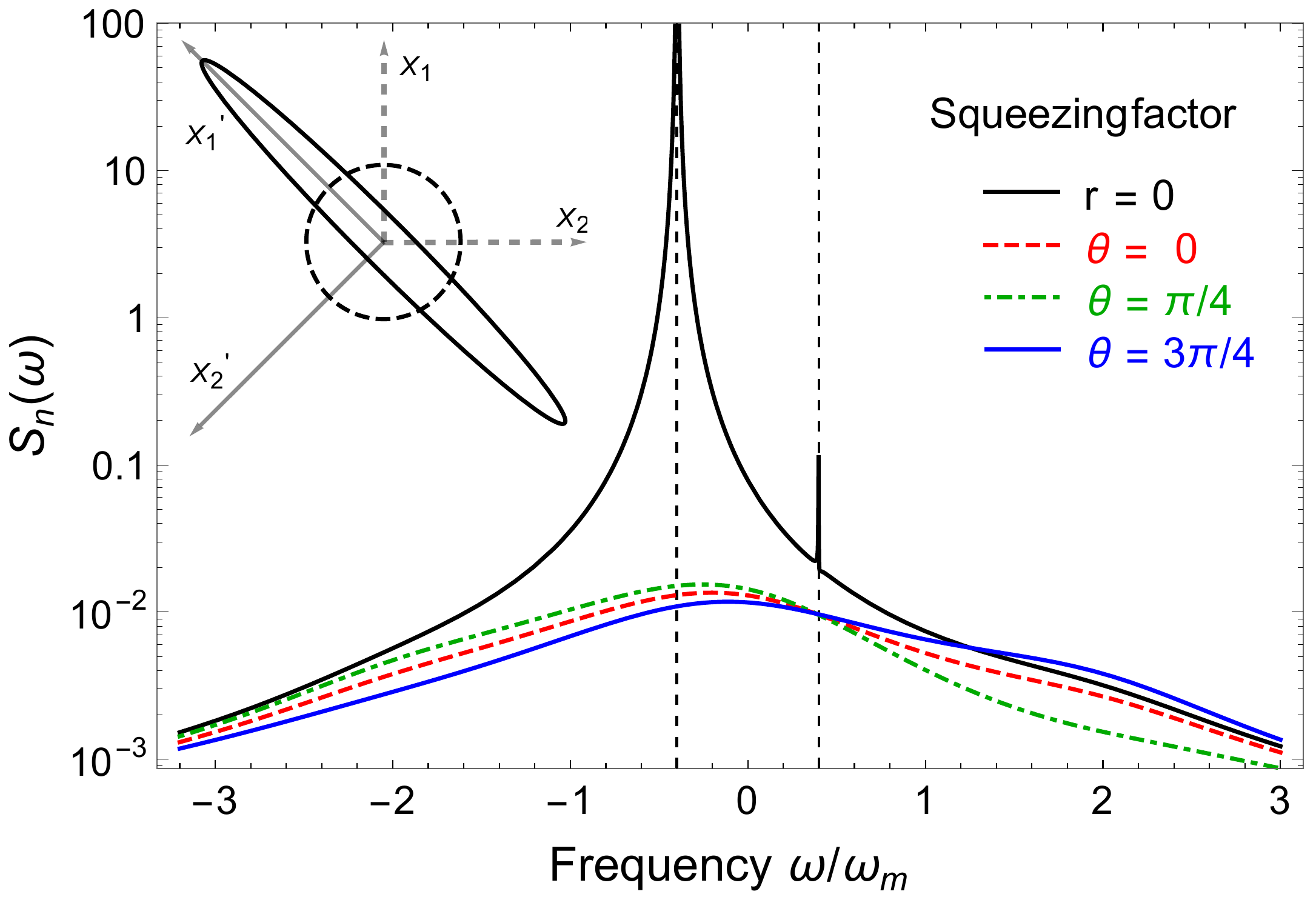}\newline
\caption{The phonon-number spectral curves of the mechanical resonator without squeezing
(black line for $r=0$) and under squeezing with a fixed squeeze factor of $r=1.2$ for different
squeezing angles $\theta$. Inset: the noise error ellipses of the two quadratures for $r=0$
(dashed circle) and for $r=1.2$ (solid ellipse). The optomechanical coupling rate $g$ is improved
to $0.01$, and the other parameters are the same as in Fig.\ref{figure3}.}
\label{figure4}
\end{figure}

Another interesting property of squeezing-based cooling is that the cooling will depend on the
squeezing angle $\theta$ when the system enters the strong-coupling regime of $g\sim \kappa$
under a higher degree of squeezing. This dependence arises because the light mode couples to the
asymmetric quadratures, whose squeezing directions are modified by the angle $\theta$. From the
perspective of the squeezing picture, the Hamiltonian given in (\ref{Heff}) will be
$\hat{S}^{\dag}(\xi)\hat{H}_{eff}\hat{S}(\xi)$, and the third term of Eq.(\ref{Heff}), which
describes  the optical coupling of the mode $\hat{a}$ to the mechanical quadrature of $\hat{X}$
defined by $\hat{b}=(\hat{X}+i\hat{P})/2$, becomes $(\cosh r-\cos 2\theta \sinh r) \hat{X}+(\sin 2\theta \sinh r)\hat{P}$.
This expression clearly shows a $\theta$ dependence of the coupling of the quadrature to the cavity mode.
Fig.\ref{figure4} shows how the phonon spectrum changes with respect to the squeezing angle $\theta$
when the squeeze factor $r$ is fixed.
The black curve is the phonon spectrum without squeezing ($r=0$), and the other curves represent
the phonon spectra under a fixed squeeze factor of $r/\omega_m=1.2$ and different squeezing angles.
The results verify the $\theta$ dependence of the cooling rate, and an optimal case arises when the
cooling laser is coupled directly to the ``hottest" quadrature with the maximum fluctuation. Both
mechanical resonant peaks at $\pm r_c/\omega_m$ appearing in Fig.\ref{figure4} (indicated by vertical
dashed lines) are due to an increase in the coupling $g$. The weak phonon resonant peak at $+r_c/\omega_m$
will be suppressed by the squeezing effect even in the strong-coupling regime. Moreover, a phonon spectrum
with a richer structure within the resolved-sideband limit or in the strong-coupling regime still exhibits
a similar squeeze-induced enhancement of the cooling effect \cite{SM}.

In summary, through direct numerical calculations based on a simple physical picture, we have revealed
a deep cooling scheme for an optomechanical resonator in which the cooling performance is dramatically
enhanced by the squeezing effect induced by parametric amplification. We have demonstrated that by increasing
the squeeze factor $r$, we can effectively reduce the area under the phonon-number spectral curve, thereby
extracting significant ``heat'' from the squeezed motion of a mechanical oscillator, to reach an effective
temperature below the quantum shot noise.
The idea proposed here is that one quadrature of the mechanical mode can be ``heated" by squeezing the other
to improve the cooling capacity by coupling the cooling laser directly to the ``heated" quadrature, allowing
the ``hotter" phonons to be quickly taken away by the leakage of photons from a bad cavity. This
method can be used to rapidly cool the mechanical motion down to its quantum ground state, beyond the standard
quantum limit. The resulting high cooling efficiency is of interest for many quantum applications, such as quantum
precision measurement or quantum sensors and rapid state initialization for quantum processing. We believe that this
squeezing-based cooling scheme can serve as a universal technique for facilitating the development of quantum
technology for use in macroscopic solid-state systems.

\begin{acknowledgements}
We thank Keye Zhang and the other active fellows in our laboratory for valuable discussions.
This work was supported by the National Natural Science Foundation of China (Grant No. 11447025)
and the Scientific Research Foundation for Returned Overseas Chinese Scholars of the
State Education Ministry.
\end{acknowledgements}


\begin{thebibliography}{99}
\bibitem{temperature} The relation between the fluctuation (energy variance
$\Delta E$) and temperature ($T$) of a classical system in equilibrium with a bath
can be expressed, according to the fluctuation-dissipation theorem,
as $\Delta E=\mu k_B T$, where $\mu$ is the mobility for the energy damping
to the equilibrium bath (e.g., the energy fluctuation in a canonical ensemble satisfies
$\Delta E\propto k_B T$). However, for a well-isolated quantum system, the relation
is ambiguous because of the variety of temperature definitions
beyond that with respect to a bath. Nevertheless, the fluctuation-dissipation theorem
is assumed to be valid in the quantum regime \cite{Gring}, and thus, we use
the word ``heated'' in the sense of an effective temperature $T_{eff}$ related to the
energy fluctuations of the system.
\end{thebibliography}
\end{document}